\documentclass[11pt]{article}
\usepackage{jcappub}
\usepackage{amsmath,amssymb,amsbsy}

\begin{document}

\begin{flushright}
{YITP17-83, IPMU17-0111}
\end{flushright}

\title{A Class of Minimally Modified Gravity Theories}

\author[a]{Chunshan Lin}

\affiliation[a]{Center for Gravitational Physics, Yukawa Institute for Theoretical Physics, Kyoto University, Kyoto 606-8502, Japan}

\author[a,b]{~~Shinji Mukohyama}

\affiliation[b]{Kavli Institute for the Physics and Mathematics of the Universe (WPI), The University of Tokyo Institutes for Advanced Study, The University of Tokyo, Kashiwa, Chiba 277-8583, Japan}

\abstract{
We investigate the Hamiltonian structure of a class of gravitational theories whose actions are linear in the lapse function. We derive the necessary and sufficient condition for a theory in this class to have two or less local physical degrees of freedom. As an application we then find several concrete examples of modified gravity theories in which the total number of local physical degrees of freedom in the gravity sector is two. 
}

\maketitle

\section{Introduction}

Searching for self-consistent extensions of general relativity is well motivated by physics at both ultra-high and ultra-low energy scales. The modification at high scale may provide a possible candidate for the unified quantum theory of gravitation. Needless to say, superstring theory is one of such examples. Horava-Lifshitz gravity~\cite{Horava:2009uw,Mukohyama:2010xz}, which draws the lessons from the anisotropic scaling of space and time in the physics of condensed matter, is another example as it is free from Ostrogradsky ghost and renormalizable~\cite{Barvinsky:2015kil}. At cosmological scale, the dark energy and dark matter problems still remain unsolved and leave open the possibility of modified gravity at large scale. One popular approach to the dark energy problem is to introduce new degrees of freedom, belonging to the gravitational sector. These new degrees of freedom might speed up the expansion of the universe. Quintessence~\cite{Ratra:1987rm,Caldwell:1997ii}, ghost condensate~\cite{ArkaniHamed:2003uy},  massive gravity~\cite{FPtheory,deRham:2010kj,Gumrukcuoglu:2011ew} are such examples.

Modifications of Einstein gravity often give rise to additional degree(s) of freedom. Particularly, if the additional degree(s) is/are due to higher derivatives,  the theory generally suffers from Ostrogradsky ghost instability \cite{Ostrogradsky}\cite{Woodard:2006nt}.  The most general scalar-tensor theory which gives rise to second order equations of motion is the Horndeski theory\cite{Horndeski} and it was rediscovered recently in the context of extensions of the so-called Galileon theory \cite{Nicolis:2008in,Deffayet:2009wt,Deffayet:2011gz}. It turned out that Horndeski's theory is not the most general scalar-tensor theory with only three local physical degrees of freedom. Recently, several examples beyond the Horndeski theory have been spotted in the literature and shown to be free from the Ostrogradsky ghost \cite{Zumalacarregui:2013pma,Gleyzes:2014dya,Gao:2014soa,Lin:2014jga,Gao:2014fra,Langlois:2015cwa,Langlois:2015skt,BenAchour:2016fzp}.

In the case of scalar tensor theories in the so called unitary gauge in which the time coordinate is chosen to agree with a fixed monotonic function of the scalar field, the additional scalar degree, which we call a scalar graviton, is nothing but a Nambu-Goldstone boson associated with the broken temporal diffeomorphism, and the action is non-linear in the lapse function. It is then intriguing to ask what if the action of the theory is linear in the lapse function, while the temporal diffeomorphism is still broken. Since the action is assumed to be linear in the lapse function here, we would naively expect that the Hamiltonian constraint eliminates the longitudinal polarization of graviton, instead of fixing the lapse function itself. Thus it might be tempting to expect that there are only two local physical degrees of freedom in the gravity sector.  However, as we will show in this paper, generally this type of theories have odd dimensional phase space at each point and thus they are not self-consistent. 

Therefore, one may wonder whether general relativity is the unique theory with two local physical degrees of freedom within the class of theories considered here. In other words, can we find some different theory with only two local physical degrees of freedom which is as good as general relativity in the sense that all constraints are first class and the structure of the theory at low energies is thus expected to be stable against quantum corrections? According to Lovelock's theorem \cite{lovelock1,lovelock2}, the only possibility in $4$-dimensions is the Einstein gravity if we impose the space-time diffeomorphism invariance in the first place. It is thus very interesting to see how Lovelock's theorem may be evaded if only the spatial diffeomorphism invariance is imposed. In this paper, for the first time, we derive the self-consistency condition for the class of theories whose actions are linear in the lapse function. By solving this consistency condition, we find several examples of modified gravity theories with two local physical degrees of freedom.

The rest of this paper is organized as follows, in section \ref{sccond}, we write down a class of actions linear in the lapse function and derive the self-consistency condition. In section \ref{theories}, we work out several solutions to the self-consistency condition. We conclude the paper in section \ref{conclusion}.

\section{A self-consistency condition}\label{sccond}

We consider a class of ($3+1$)-dimensional theories that are invariant under the spatial diffeomorphism, 
\begin{eqnarray}
 x^i\to x^i+\xi^i(t,\bf{x})\,,
\end{eqnarray}
where $x^i$ ($i=1,2,3$) are spatial coordinates. Since we do not require the invariance under the temporal diffeomorphism, it is convenient to adopt the ADM decomposition of the $4$-dimensional metric, 
\begin{eqnarray}
ds^2=-N^2dt^2+h_{ij}\left(dx^i+N^idt\right)\left(dx^j+N^jdt\right)\,,
\end{eqnarray}
where $N$ is the lapse function, $N^i$ is the shift vector and $h_{ij}$ is the $3$-dimensional spatial metric. In the present paper we are interested in those theories whose actions are linear in the lapse function and are of the form
\begin{eqnarray}\label{genst1}
 S=\int dt d^3x\sqrt{h}N F\left(K_{ij},R_{ij},\nabla_i,h^{ij}, t\right)\,,
\end{eqnarray}
where $K_{ij}=(\partial_th_{ij}-\nabla_iN_j-\nabla_jN_j)/(2N)$ and $R_{ij}$ are the extrinsic curvature and the Ricci tensor of the constant-$t$ hypersurfaces, respectively, and $\nabla_{i}$ is the covariant derivative compatible with the induced metric $h_{ij}$. All indices in each term of $F$ in the action must be contracted via the induced metric $h_{ij}$ and its inverse $h^{ij}$ to form a spatial scalar. As always, it is possible to recover the $4$-dimensional general covariance by introducing a scalar field. From the point of view of such a covariant description of the same theory, the action (\ref{genst1}) in terms of ($N$, $N^i$, $h_{ij}$) without the scalar field is nothing but the action in the so called unitary gauge, in which the time coordinate is chosen to agree with a fixed monotonic function of the scalar field. In the present paper, we shall study the class of theories in the unitary gauge.

We exclude the case where the action contains mixed space-time derivative terms, i.e. terms that contain spatial derivatives of the extrinsic curvature such as $h^{il}h^{jm}h^{kn}\nabla_iK_{jk}\nabla_lK_{mn}$, because the Hamiltonian structure in this case is quite different from the case without mixed derivative terms. (With mixed derivative terms, $\Phi^{ij}$ in (\ref{secons3}) below would include spatial derivatives of $\ln N$. See Appendix \ref{mixder} for details.) We further assume that
\begin{equation}
 \det\left(\frac{\partial^2 F}{\partial K_{ij}\partial K_{kl}}\right) \ne 0\,,
  \label{eqn:invertibility-pre}
\end{equation}
so that the relation between $K_{ij}$ and the momenta conjugate to $h_{ij}$ is invertible.

To perform the Hamiltonian analysis by means of Dirac's method, we introduce two auxiliary tensor fields $Q_{ij}$ and $\upsilon^{ij}$, and an equivalent action is written as 
\begin{eqnarray}\label{actionvQ}
 S = \int d^4x\mathcal{L}\,, \quad
  \mathcal{L} = \sqrt{h}N\left[F\left(Q_{ij},R_{ij},\nabla_i,h^{ij}, t\right)+\upsilon^{ij}\left(Q_{ij}-K_{ij}\right)\right]\,.
\end{eqnarray} 
The last term in this action enforces a constraint setting the auxiliary tensor field $Q_{ij}$ to the extrinsic curvature $K_{ij}$ through the equation of motion for $\upsilon^{ij}$. After imposing this constraint, we recover the original action (\ref{genst1}). The condition (\ref{eqn:invertibility-pre}) translates to 
\begin{equation}
 \det\left(\frac{\partial^2 F}{\partial Q_{ij}\partial Q_{kl}}\right) \ne 0\,.
  \label{eqn:invertibility}
\end{equation}
Considering the action (\ref{actionvQ}), the momenta conjugate to ($h_{ij}$, $N$, $N^i$, $Q_{ij}$, $\upsilon^{ij}$) are calculated respectively as 
\begin{eqnarray}\label{momenta}
\pi^{ij}&=&\frac{\partial\mathcal{L}}{\partial\dot{h}_{ij}}=-\frac{1}{2}\sqrt{h}\upsilon^{ij},\qquad \pi_N=\frac{\partial\mathcal{L}}{\partial\dot{N}}=0,\qquad\pi_i=\frac{\partial\mathcal{L}}{\partial\dot{N}^{i}}=0\,,\nonumber\\
P^{ij}&=&\frac{\partial\mathcal{L}}{\partial\dot{Q}_{ij}}=0,\qquad U_{ij}=\frac{\partial\mathcal{L}}{\partial\dot{\upsilon}^{ij}}=0\,.
\end{eqnarray}
The Hamiltonian then reads
\begin{eqnarray}
H&=&\int d^3x\left[ \pi^{ij}\dot{h}_{ij}-\mathcal{L}+\lambda_N\pi_N+\lambda^i\pi_i
+\chi_{ij}P^{ij}+\varphi^{ij}U_{ij}+\lambda_{ij}\Psi^{ij}\right]\nonumber\\
&=&\int d^3x\left[N\mathcal{C}+N^i\mathcal{H}_i+\lambda_N\pi_N+\lambda^i\pi_i+\chi_{ij}P^{ij}+\varphi^{ij}U_{ij}+\lambda_{ij}\Psi^{ij}\right]\,,
\end{eqnarray}
where ($\lambda_N$, $\lambda^i$, $\chi_{ij}$, $\varphi^{ij}$, $\lambda_{ij}$) are Lagrange multipliers, 
\begin{eqnarray}\label{momconst}
\mathcal{C}& \equiv &-\sqrt{h}\left[F\left(Q_{ij},R_{ij},\nabla_i,h^{ij}, t\right)+\upsilon^{ij}Q_{ij}\right]\,,\nonumber\\
\mathcal{H}_i& \equiv &\sqrt{h}\nabla_j\upsilon_{~i}^{j}\,,\nonumber\\
\Psi^{ij}&\equiv&\pi^{ij}+\frac{1}{2}\sqrt{h}\upsilon^{ij}\,,
\end{eqnarray}
and we have used the first equation in (\ref{momenta}) (or equivalently we have redefined the Lagrange multiplier $\lambda_{ij}$) to eliminate the ``velocity'' $\dot{h}_{ij}$ in the Hamiltonian. There are $22$ primary constraints,
\begin{eqnarray}
\pi_N\approx0,\qquad\pi_i\approx0,\qquad P^{ij}\approx0\,,
\qquad U_{ij}\approx0,\qquad\Psi^{ij}\approx0\,.
\end{eqnarray}
To be consistent, these $22$ primary constraints must be preserved by time evolution of the system. The consistency conditions then give the following $10$ secondary constraints,
\begin{eqnarray}\label{secons}
0 & \approx & \frac{d\pi_N}{dt} = \{\pi_N,H\} = -\mathcal{C}\,,\label{secons1}\\
0 & \approx & \frac{d\pi_i}{dt} = \{\pi_i,H\}=-\mathcal{H}_i\,,\label{secons2}\\
0 & \approx & \frac{dP^{ij}}{dt} = \{P^{ij},H\} = N\Phi^{ij}\,,\label{secons3}
\end{eqnarray}
where $\{\cdots,\cdots\}$ denotes the Poisson bracket and
\begin{equation}
 \Phi^{ij} \equiv \sqrt{h}\left(\frac{\partial F}{\partial Q_{ij}}+v^{ij}\right)\,.
  \label{eqn:defPhiij}
\end{equation}
For a scalar density (or a scalar) $\mathcal{O}$ and a vector density (or a vector) $\mathcal{O}_i$, we define the following useful notations, 
\begin{eqnarray}
\bar{\mathcal{O}}[\lambda]\equiv \int  d^3x\lambda \mathcal{O}\,,\qquad \bar{\mathcal{O}}_i[\lambda^i]\equiv \int  d^3x\lambda^i \mathcal{O}_i\,,
\end{eqnarray}
where $\lambda$ and $\lambda_i$ are test smooth functions that behave as a scalar (or a scalar density) and a vector (or a vector density), respectively. The consistency conditions for the rest of $12$ primary constraints $U_{ij}\approx0,~\Psi^{ij}\approx0$ only fix the Lagrange multipliers in front of them. Let us collect all of primary and secondary constraints in the total Hamiltonian and treat all of them on the same footing, 
\begin{eqnarray}
H_{\rm tot}=\int d^3x\left[\lambda_c\mathcal{C}+\tilde{N}^i\mathcal{H}_i+\lambda_N\pi_N+\lambda^i\pi_i+\chi_{ij}P^{ij}+\varphi^{ij}U_{ij}+\lambda_{ij}\Psi^{ij}+\phi_{ij}\Phi^{ij}\right]\,,
\end{eqnarray}
where ($\lambda_c$, $\tilde{N}^i$, $\lambda_N$, $\lambda^i$, $\chi_{ij}$, $\varphi^{ij}$, $\lambda_{ij}$, $\phi_{ij}$) are Lagrange multipliers and we have absorbed the lapse function and the shift vector into $\lambda_c$ and $\tilde{N}^i$, respectively.

It is easy to check that $\pi_i\approx0$ are first class and they eliminate the conjugate canonical pairs $\left(N^i,\pi_i\right)$. Moreover, due to the spatial diffeomorphism invariance of the theory, the extended momentum constraints,
 \begin{eqnarray}\label{emc}
  \mathcal{H}^E_{i}&=&-2\sqrt{h}\nabla_j\left(\frac{\pi^{j}_{~i}}{\sqrt{h}}\right)
   +P^{jk}\nabla_iQ_{jk}-2\sqrt{h}\nabla_k\left(\frac{P^{jk}}{\sqrt{h}}Q_{ij}\right)
   \nonumber\\
  & &
   +U_{jk}\nabla_i\upsilon^{jk}+2\sqrt{h}\nabla_k\left(\upsilon^{jk}\frac{U_{ij}}{\sqrt{h}}\right)+\pi_N\partial_iN\,,
\end{eqnarray}
are also first class. As one can easily guess from 
\begin{equation}
 \bar{\mathcal{H}}^E_i[\lambda^i] = \int d^3x \left[ \pi^{ij}\mathcal{L}_{\lambda}h_{ij} + P^{ij}\mathcal{L}_{\lambda}Q_{ij} + U_{ij}\mathcal{L}_{\lambda}\upsilon^{ij} + \pi_N\mathcal{L}_{\lambda}N\right] + \mbox{boundary terms}\,,
\end{equation}
where $\mathcal{L}_{\lambda}$ is the Lie derivative along the vector $\lambda^i$, they are the generators of spatial coordinate transformation (see a proof in Ref.  \cite{Saitou:2016lvb} as well as another independent proof in Ref. \cite{Lin:2017utd}).

Now let us check the property of $\pi_N\approx0$. We have 
\begin{eqnarray}
\{\bar{\pi}_{N}[\lambda],\bar{\pi}_{N}[g]\}\approx0\,, \qquad\{\bar{\pi}_{N}[\lambda],\bar{\mathcal{C}}[g]\}\approx0\,,\qquad\{\bar{\pi}_{N}[\lambda],\bar{U}[\varphi^{ij}]\}\approx0\,,\nonumber\\
\{\bar{\pi}_{N}[\lambda],\bar{\mathcal{H}}_{i}[f]\}\approx0\,,\qquad\{\bar{\pi}_{N}[\lambda],\bar{\Psi}[\lambda_{ij}]\}\approx0\,,\qquad \{\bar{\pi}_{N}[\lambda],\bar{\pi}_{i}[\lambda^i]\}\approx0\,,\nonumber\\
\{\bar{\pi}_{N}[\lambda],\bar{P}[\chi_{ij}]\}\approx0\,, \qquad \{\bar{\pi}_{N}[\lambda],\bar{\Phi}[\phi_{ij}]\}\approx0\,,
\end{eqnarray}
and all vanish weakly. Therefore $\pi_N\approx 0$ is a first class constraint and eliminates the conjugate canonical pair $\left(N,\pi_N\right)$.

We now know that $\pi_i$, $\mathcal{H}^E_{i}$ and $\pi_N$ are first class. The complete set of other independent constraints is 
\begin{eqnarray}
\mathcal{C}\approx0\,,\qquad P^{ij}\approx0\,,\qquad U_{ij}\approx0\,,\qquad\Phi^{ij}\approx0\,,\qquad \Psi^{ij}\approx0\,. 
\end{eqnarray}
In total, we have $25$ remaining constraints at each point. Let us denote these $25$ constraints at each point as $\phi_{a}\equiv\left(\mathcal{C},P^{ij}, U_{ij},\Phi^{ij},\Psi^{ij}\right)$, where $a=1,\cdots,25$ and $(ij)=(11), (22), (33), (12), (23), (31)$. The dimension of the physical phase space crucially depends on the determinant of the infinite dimensional matrix made of Poisson brackets 
\begin{eqnarray}\label{bigM}
M_{ab}(x,y)\equiv\{\phi_{a}(x),\phi_{b}(y)\}\approx
\left(\begin{array}{ccccc}
0 & 0_{6}^{T} & u_{1}^{T} & 0_{6}^{T} & \hat{u}_{2}^{T}\\
0_{6} & 0_{6,6} & 0_{6,6} & A_{1} & 0_{6,6}\\
-u_{1} & 0_{6,6} & 0_{6,6} & a\mathbf{1}_{6,6} & b\mathbf{1}_{6,6}\\
0_{6} & -A_{1}^{T} & -a\mathbf{1}_{6,6} & 0_{6,6} & \hat{A}_{2}\\
-\hat{u}_{2} & 0_{6,6} & -b\mathbf{1}_{6,6} & -\hat{A}_{2} & A_{3}
\end{array}\right),
\end{eqnarray}
where $0$ on the right hand side actually represents zero multiplied by $\delta^3(x-y)$ and thus is an infinite dimensional matrix by itself, $0_{6}$ is a $6$-entry vector whose components present zero multiplied by $\delta^3(x-y)$, $0_{6,6}$ is a $6\times6$ matrix whose components represent zero multiplied by $\delta^3(x-y)$, $\mathbf{1}_{6,6}$ is a $6\times6$ unit matrix multiplied by $\delta^3(x-y)$, $a$ and $b$ are proportional to $\delta^3(x-y)$, $u_1$ is a $6$-entry vector whose components are proportional to $\delta^3(x-y)$, $\hat{u}_2$ is a $6$-entry vector whose components are linear combinations of $\delta^3(x-y)$ and its derivatives, $A_1$ and $A_3$ are $6\times 6$ matrices that are proportional to $\delta^3(x-y)$, $\hat{A}_2$ is a $6\times 6$ matrix whose components are linear combinations of $\delta^3(x-y)$ and its derivatives. See (\ref{eqn:explicitcomponents}) below for explicit forms of ($a$, $b$, $u_1$, $\hat{u}_2$, $A_1$, $\hat{A}_2$, $A_3$).

If $\mathrm{Det} M_{ab}(x,y)\neq0$, all of $\phi_{a}\equiv\left(\mathcal{C},P^{ij}, U_{ij},\Phi^{ij},\Psi^{ij}\right)$ are second class. The algebra in this case closes here since the consistency condition of all these $25$ constraints only fixes the Lagrange multipliers in front of them. At each point there are $22$ conjugate pairs and thus $44$ degrees in the phase space, i.e. $\left(h_{ij},\pi^{ij}\right),~\left(Q_{ij},P^{ij}\right),~\left(\upsilon^{ij},U_{ij}\right),~\left(N, \pi_N\right)$ and $\left(N^i,\pi_i\right)$. The $7$ first class constraints $\pi_N\approx0$, $\pi_i\approx0$ and $\mathcal{H}^E_i\approx0$ eliminate 14 phase space degrees. The remaining $25$ second class constraints $\phi_{a}\equiv\left(\mathcal{C},P^{ij}, U_{ij},\Phi^{ij},\Psi^{ij}\right) $ eliminate 25 phase space degrees. At the end the reduced phase space dimension at each point would be $44-14-25=5$, and it is odd! Generally, odd dimensional phase space at each point leads to inconsistency. (Needless to say, the total number of the phase space dimensions is infinite since we are dealing with a field theory.) For instance, a naive (and wrong) non-projectable extension of Horava-Lifshitz gravity is inconsistent in this sense  \cite{Henneaux:2009zb} while the correct non-projectable extension does not have this problem \cite{Blas:2009yd}.

Therefore, we need to demand that $\mathrm{Det} M_{ab}(x,y)\approx0$ to ensure the consistency of the theory. In other words, we need to demand that the infinite-dimensional matrix $M_{ab}(x,y)$ has an eigenvector with a vanishing eigenvalue, which can be represented formally as a $25$-entry vector field $\textbf{v}$ satisfying 
\begin{eqnarray}
 (\textbf{M}\cdot \textbf{v})_a(x) = \int d^3y \sum_{b=1}^{25}M_{ab}(x,y)\upsilon^b(y) \approx0\,, \quad \mbox{for } {}^{\forall}a\,,\ {}^{\forall}x\,.
  \label{eqn:def-zeroeigenvector}
\end{eqnarray}
Decomposing components of $\textbf{v}$ as $\textbf{v}=\left(\alpha,v_1,v_2,v_3,v_4\right)^T$, where $\alpha$ is a scalar field and $v_{1,2,3,4}$ are $6$-entry vector fields, the condition (\ref{eqn:def-zeroeigenvector}) is rewritten as the following set of conditions,
\begin{eqnarray}
\int d^3y \left[u_1^T\upsilon_2(y)+\hat{u}_2^T\upsilon_4(y)\right]\approx0\,, \quad \mbox{for } {}^{\forall}x\,,\nonumber\\
\int d^3y \left[\hat{A}_1\upsilon_3(y)\right]\approx0\,, \quad \mbox{for } {}^{\forall}x\,,\nonumber\\
\int d^3y \left[-u_1\alpha(y)+a\upsilon_3(y)+b\upsilon_4(y)\right]\approx0\,, \quad \mbox{for } {}^{\forall}x\,,\nonumber\\
\int d^3y \left[-A_1^T\upsilon_1(y)-a\upsilon_2(y)+\hat{A}_2\upsilon_4(y)\right]\approx0\,, \quad \mbox{for } {}^{\forall}x\,,\nonumber\\
\int d^3y \left[-\hat{u}_2\alpha(y)-b\upsilon_2(y)-\hat{A}_2\upsilon_3(y)+A_3\upsilon_4(y)\right]\approx0\,, \quad \mbox{for } {}^{\forall}x\,. \label{eqn:consistency-condition-decomposed}
\end{eqnarray}
Note that the left hand side of each equation depends on $x$. If one finds a non-vanishing solution to the set of equations (\ref{eqn:consistency-condition-decomposed}) then the liner combination
\begin{equation}
 \int d^3x \sum_a\upsilon^a(x)\phi_a(x) \label{eqn:1stclass}
\end{equation}
is first class, provided that
\begin{equation}
 \int d^3x \sum_a\upsilon^a(x)\frac{\partial \phi_a(x)}{\partial t}\label{eqn:intvphi}
\end{equation}
weakly vanishes. If $\textbf{v}=\left(\alpha,v_1,v_2,v_3,v_4\right)^T$ satisfies (\ref{eqn:consistency-condition-decomposed}) and if (\ref{eqn:intvphi}) does not vanish weakly, then (\ref{eqn:intvphi})$\approx 0$ will be a tertiary constraint.

Using the explicit expressions of the components of $M_{ab}(x,y)$,
\begin{eqnarray}
 & & a = 2b = -\sqrt{h}\delta^3(x-y)\,,\quad
 u_1^T = -\sqrt{h}Q_{ij}\delta^3(x-y)\,, \quad
 \hat{u}_2^T = \frac{\delta \mathcal{C}(x)}{\delta h_{ij}(y)}\,,\nonumber\\
& &  A_1^T = -\sqrt{h}\frac{\partial^2F}{\partial Q_{ij}\partial Q_{kl}}\delta^3(x-y)\,,\quad
 \hat{A}_2^T = \frac{\delta \Phi^{ij}(x)}{\delta h_{kl}(y)}\,,\nonumber\\
& &  A_3^T = \frac{\sqrt{h}}{4}(\upsilon^{ij}h^{kl}-\upsilon^{kl}h^{ij})\delta^3(x-y)\,, \label{eqn:explicitcomponents}
\end{eqnarray}
it is straightforward to show that
\begin{equation}\label{cond1pre}
\int d^3x \left[ \frac{\delta \bar{\mathcal{C}}[\beta]}{\delta h_{ij}(x)}Q_{ij}(x)\alpha(x)-\frac{\delta \bar{\mathcal{C}}[\alpha]}{\delta h_{ij}(x)}Q_{ij}(x)\beta(x)\right]\approx0\,, \quad \mbox{for } {}^{\forall}\beta(x)\,.
\end{equation}
Once this condition is fulfilled by the component $\alpha$, other components $v_{1,2,3,4}$ are uniquely expressed in terms of $\alpha$. In particular (\ref{eqn:invertibility}) implies that $v_3=0$ and that (\ref{eqn:1stclass}) and (\ref{eqn:intvphi}) do not contain $\Phi^{ij}$ and $\partial\Phi^{ij}/\partial t$, respectively. The condition (\ref{cond1pre}) is trivially satisfied if $\alpha(x)$ vanishes everywhere, but in this case $\textbf{v}=0$, (\ref{eqn:1stclass}) and (\ref{eqn:intvphi}) vanish, and thus there is neither associated first-class constraint nor associated tertiary constraint. We therefore need to demand that there exists a function $\alpha(x)$ that does not vanish everywhere and that satisfies (\ref{cond1pre}). However, since (\ref{eqn:1stclass}) and (\ref{eqn:intvphi}) are not functions but integrals of functions over the space, having one such function $\alpha(x)$ is not enough. In order to have either a first class constraint or a tertiary constraint at every point of the space, (\ref{cond1pre}) needs to be satisfied by arbitrary $\alpha(x)$. In summary, the necessary and sufficient condition for a theory in the class of theories considered in this section to have two or less local physical degrees of freedom is 
\begin{eqnarray}\label{cond1pre2}
\int d^3x\left[ \frac{\delta \bar{\mathcal{C}}[\beta]}{\delta h_{ij}(x)}Q_{ij}(x)\alpha(x)-\frac{\delta \bar{\mathcal{C}}[\alpha]}{\delta h_{ij}(x)}Q_{ij}(x)\beta(x)\right]\approx0\,, \quad \mbox{for } {}^{\forall}\alpha(x)\,,\ {}^{\forall}\beta(x)\,.
\end{eqnarray}
This condition can be rewritten as
\begin{eqnarray}
 & & \int d^3x\sqrt{h}
  \left\{\nabla^j
   \left(\frac{1}{\sqrt{h}}\frac{\tilde{\delta}\bar{F}[\sqrt{h}\alpha]}{\tilde{\delta} R_{kl}(x)}
    \right)
   \nabla^i\left[\left(Q_{jl}h_{ik}-\frac{1}{2}Q_{kl}h_{ij}-\frac{1}{2}Qh_{ik}h_{jl}\right)\beta\right] -(\alpha\leftrightarrow\beta)\right\}\approx 0\,,\nonumber\\
 & & \hspace{8cm}
  \mbox{for } {}^{\forall}\alpha(x)\,,\ {}^{\forall}\beta(x)\,,
  \label{cond1}
\end{eqnarray}
where $Q=h^{ij}Q_{ij}$. Here, $\tilde{\delta}/\tilde{\delta}R_{ij}$ is the functional derivative with respect to $R_{ij}$ when ($R_{ij}$, $h_{ij}$, $\alpha$, $\beta$) are considered as independent variables. The condition (\ref{cond1}), or equivalently (\ref{cond1pre2}), is the self-consistency condition for the class of theories under consideration. This is the main result of the present paper.

Under the condition (\ref{cond1}), or equivalently (\ref{cond1pre2}), we define a local function $\mathcal{C}^E(x)$ so that 
\begin{equation}
 \int d^3x\sum_a v^a(x)\phi_a(x) = \bar{\mathcal{C}}^E[\alpha] + \mbox{boundary terms} \label{eqn:def-CE}
\end{equation}
for any $\textbf{v}=\left(\alpha,v_1,v_2,v_3,v_4\right)^T$ satisfying the set of conditions (\ref{eqn:consistency-condition-decomposed}). By definition, $\mathcal{C}^E(x)$ is a linear combination of constraints and has an $O(1)$ overlap with the Hamiltonian constraint $\mathcal{C}$. We thus call $\mathcal{C}^E$ the extended Hamiltonian constraint. As already stated, the assumption (\ref{eqn:invertibility}) implies $v_3=0$ for ${}^{\forall}\alpha$ and thus $\mathcal{C}^E$ does not contain $\Phi^{ij}$. If $\partial\mathcal{C}^E/\partial t\approx 0$ then $\mathcal{C}^E$ is first class. If $\partial\mathcal{C}^E/\partial t$ does not vanish weakly then $\partial\mathcal{C}^E/\partial t\approx 0$ should be imposed as a tertiary constraint. Since $\mathcal{C}^E$ does not contain $\Phi^{ij}$ and all other constraints except $\mathcal{C}$ are time-independent, we have
\begin{equation}
 \frac{\partial \mathcal{C}^E}{\partial t} \approx \frac{\partial \mathcal{C}}{\partial t}\,.
\end{equation}
The equality holds only weakly in general since some of the coefficients of other constraints may be time-dependent, given that F contains time as one of its arguments.

For $F=F\left(Q_{ij},R_{ij},h^{ij},t\right)$, the condition (\ref{cond1}) simplifies to
\begin{eqnarray}
& &\int\sqrt{h}\left(\alpha\nabla^i\beta-\beta\nabla^i\alpha\right)\left[-\frac{\partial F}{\partial R_{kl}}\nabla^j\left(Q_{il}h_{jk}-\frac{1}{2}Q_{kl}h_{ij}-\frac{1}{2}Qh_{jk}h_{il}\right)\right.\nonumber\\
 &&\quad \left.+\nabla^j\left(\frac{\partial F}{\partial R_{kl}}\right)\cdot\left(Q_{jl}h_{ik}-\frac{1}{2}Q_{kl}h_{ij}-\frac{1}{2}Qh_{ik}h_{jl}\right)\right]\approx 0\,,\quad
  \mbox{for } {}^{\forall}\alpha(x)\,,\ {}^{\forall}\beta(x)\,.
\label{cond2pre}
\end{eqnarray}
A sufficient condition to satisfy (\ref{cond2pre}) is
\begin{eqnarray}
& &-\frac{\partial F}{\partial R_{kl}}\nabla^j\left(Q_{il}h_{jk}-\frac{1}{2}Q_{kl}h_{ij}-\frac{1}{2}Qh_{jk}h_{il}\right)\nonumber\\
 &&\quad +\nabla^j\left(\frac{\partial F}{\partial R_{kl}}\right)\cdot\left(Q_{jl}h_{ik}-\frac{1}{2}Q_{kl}h_{ij}-\frac{1}{2}Qh_{ik}h_{jl}\right)\approx 0\,. 
\label{cond2}
\end{eqnarray}
For $F=F\left(Q_{ij},h^{ij},t\right)$, the condition (\ref{cond2}) is trivially satisfied but theories of this type are already excluded by experiments and observations since gravitational waves would have a vanishing propagating speed and a static mass would not gravitate. In the next section we therefore consider some simple examples in which $F$ depends on $R_{ij}$ and thus the condition (\ref{cond2}) is non-trivial.

\section{Theories with two local physical degrees of freedom}\label{theories}

In this section, as an application of the condition (\ref{cond1}), or equivalently (\ref{cond1pre2}), we seek concrete examples of modified gravity theories in which the total number of local physical degrees of freedom in the gravity sector is two. For simplicity we consider the simple case with $F=F\left(Q_{ij},R_{ij},h^{ij},t\right)$ and try to solve the self-consistency condition (\ref{cond2}) by adopting several simple ansatz. As the first example, we shall consider Einstein's general relativity. We then present several modified gravity theories that have two local physical degrees of freedom.

\subsection{General relativity}

As the first example, let us consider the simple ansatz, 
\begin{eqnarray}
 F=f_1(\mathcal{Q})+f_2(R)\,, \quad
  f'_1(\mathcal{Q}) \ne 0\,,\ f'_2(R) \ne 0\,,
\end{eqnarray}
where $f_1'$ and $f_2'$ are the derivative of $f_1$ and $f_2$, respectively, with respect to their argument, $\mathcal{Q}\equiv Q_{ij}Q^{ij}-Q^2$ and $Q=Q^i_{\ i}$. With this ansatz we try to find a solution to (\ref{cond2}), which is 
\begin{eqnarray}
 -\nabla^i\left(Q_{ij}-Qh_{ij}\right)+\left(Q_{ij}-Qh_{ij}\right)\nabla^i\ln f_2' \approx0\,. \label{eqn:cond2-ex1}
\end{eqnarray}
Since this condition is a weak equality, we need to take into account constraints. In particular, the momentum constraint (\ref{secons2}) combined with (\ref{secons3}) gives
\begin{eqnarray}
\nabla^i\left(Q_{ij}-Qh_{ij}\right)+\left(Q_{ij}-Qh_{ij}\right)\nabla^i\ln f_1'\approx0\,,
\end{eqnarray}
and makes it possible for us to rewrite (\ref{eqn:cond2-ex1}) as
\begin{eqnarray}
\partial_i\ln\left(f_1'f_2'\right)\approx0\,, \qquad\text{i.e.}\qquad f_1'(\mathcal{Q}) f'_2(R)\approx\text{constant in space.}
\end{eqnarray}
This is satisfied if both of $f_1'(\mathcal{Q})$ and $f_2'(R)$ are constant in space. In this case we have
\begin{eqnarray}\label{theory0}
 F=c_1(t) \mathcal{Q} + c_2(t)R-\Lambda(t)\,, \quad
  c_1(t)\ne 0\,, \ c_2(t)\ne 0\,.
\end{eqnarray}
This is nothing but Einstein gravity, if all coefficients $c_1$, $c_2$ and $\Lambda$ are constant and if $c_1$ and $c_2$ are positive. If $c_1$ and/or $c_2$ are/is negative then the theory exhibits ghost and/or gradient instability at scales shorter than the curvature scale.

On the other hand, if one or more of the coefficients $c_1$, $c_2$ and $\Lambda$ are time dependent, then the consistency condition requires that Hamiltonian constraint should be preserved by the time evolution, 
\begin{eqnarray}
0\approx\frac{d\mathcal{C}}{dt}=\frac{\partial \mathcal{C}}{\partial t}+\{\mathcal{C},H_{\rm tot}\}\approx\frac{\partial \mathcal{C}}{\partial t}\,. 
\end{eqnarray}
This is automatically satisfied and the extended Hamiltonian constraint $\mathcal{C}^E$ defined by (\ref{eqn:def-CE}) remains first class if
\begin{equation}
c_1c_2 = \text{constant}\,, \quad  c_1\Lambda = \text{constant}\,. \label{eqn:c1L-c2-L-const}
\end{equation}
In this case, the time-dependence of $c_1$, $c_2$ and $\Lambda$ can be removed by redefinition of the lapse function $N$ and thus the theory is equivalent to general relativity, provided that both $c_1$ and $c_2$ are positive. If (\ref{eqn:c1L-c2-L-const}) is satisfied and if $c_1$ and/or $c_2$ are/is negative then the theory exhibits ghost and/or gradient instability at scales shorter than the curvature scale. If 
\begin{equation}
c_1c_2 = \text{constant}\,, \quad  c_1\Lambda \ne \text{constant}\,,
\end{equation}
then $\partial\mathcal{C}^E/\partial t = \partial\mathcal{C}/\partial t\approx \Lambda\partial_t\ln (c_1\Lambda)$ is a fixed non-vanishing function of time and thus the theory is inconsistent. Finally, if
\begin{equation}
c_1c_2\ne \text{constant}\,, 
\end{equation}
then $0\approx\partial\mathcal{C}^E/\partial t = \partial\mathcal{C}/\partial t$ gives a tertiary constraint, whose Poisson bracket with the extended Hamiltonian constraint $\mathcal{C}^E$ does not vanish. In this case, both of the extended Hamiltonian constraint $\mathcal{C}^E\approx 0$ and the tertiary constraint $\partial\mathcal{C}^E/\partial t\approx 0$ are second class, and the scalar graviton is eliminated by the two second class constraints.

\subsection{A square root gravity}
\label{subsec:squareroot}

To find other examples, let us take the following ansatz 
\begin{eqnarray}
 F=f_1\left(\mathcal{Q}\right)f_2(R)-\Lambda(t)\,, \quad
  f'_1(\mathcal{Q}) \ne 0\,,\ f'_2(R) \ne 0\,.
\end{eqnarray}
Plugging the ansatz into (\ref{cond2}), we obtain
\begin{eqnarray}
-\nabla^i\left(Q_{ij}-Qh_{ij}\right)+\nabla^i\ln\left(f_1f_2'\right)\cdot\left(Q_{ij}-Qh_{ij}\right)\approx0\,. \label{eqn:cond2-ex2}
\end{eqnarray}
The momentum constraint $\nabla_i\left(\partial F/\partial Q_{ij}\right)\approx 0$ is 
\begin{eqnarray}
 \nabla^i\left(Q_{ij}-Qh_{ij}\right)+\nabla^i\ln\left(f_1'f_2\right)\cdot\left(Q_{ij}-Qh_{ij}\right)\approx0\,, 
\end{eqnarray}
and allows one to rewrite (\ref{eqn:cond2-ex2}) as
\begin{eqnarray}
 \partial_i\ln\left( f_1f_1' f_2f_2'\right)\approx0, \qquad\text{i.e.}\qquad f_1(\mathcal{Q})f_1'(\mathcal{Q}) f_2(R)f'_2(R)\approx\text{constant in space.}
\end{eqnarray}
This is satisfied if 
\begin{eqnarray}\label{f1f2}
 f_1^2=A(t)\mathcal{Q}+B(t)\,,\quad f_2^2=C(t)R+D(t)\,,\quad
  A(t)\ne 0\,,\ C(t)\ne 0\,.
\end{eqnarray}
where $A, B, C$ and $D$ are integration ``constants'' that may depend on time.

If all coefficients, i.e. $A$ ($\ne 0$), $B$, $C$ ($\ne 0$), $D$ and $\Lambda$, are constants, then the constraint algebra closes here. The extended Hamiltonian constraint in this case is first class, and thus there are only two local physical degrees of freedom in the gravity sector. In Appendix \ref{apsqq} we perform the Hamiltonian analysis without introducing auxiliary tensors $Q_{ij}$ and $\upsilon^{ij}$, and confirm the same results. If $B=\Lambda=0$ then the theory with constant coefficients is actually equivalent to Einstein's gravity. Indeed, this is the so-called shape dynamics description of general relativity, whose basic idea and formula were derived in Ref. \cite{shape1}. (See also Ref. \cite{Mercati:2014ama} for an introductory review of the shape dynamics.) The Baierlein-Sharp-Wheeler (BSW) action of the shape dynamics can be obtained by solving the Hamiltonian constraint of general relativity with respect to the lapse function, and then plugging the solution back into the action, See Appendix \ref{bsw} for the derivation of the BSW action\footnote{A square-root form of the Hamiltonian of general relativity was obtained in Ref. \cite{Murchadha:2012zz} by solving a constraint equation at the Hamiltonian level.}. If $B=0$ and $\Lambda\ne 0$ then the Hamiltonian constraint becomes $\mathcal{C}\approx \sqrt{h}\Lambda\ne 0$ and thus the theory is inconsistent. On the other hand, for $B\neq 0$, the theory with constant coefficients does not have this problem, and is different from the shape dynamics description of general relativity since one can not obtain this action by solving the Hamiltonian constraint of general relativity.

If these coefficients are time dependent, the consistency of the Hamiltonian constraint (\ref{secons1}) with the time evolution requires that 
\begin{equation}
 0 \approx \frac{d\mathcal{C}^E}{dt}
  = \frac{\partial \mathcal{C}^E}{\partial t}+\{\mathcal{C}^E,H_{\rm tot}\}
  \approx \frac{\partial \mathcal{C}^E}{\partial t}
  \approx \frac{\partial \mathcal{C}}{\partial t} 
  = -\sqrt{h}\frac{\partial F}{\partial t}\,. \label{eqn:dFdt=0_sqrt}
\end{equation}
Without loss of generality, we can assume that $A\mathcal{Q}+B\geq 0$~\footnote{If $A\mathcal{Q}+B< 0$ then one can flip the sign of $A\mathcal{Q}+B$ without changing the action by the replacement $A\to -A$, $B\to -B$, $C\to -C$, $D\to -D$.}. A combination of the Hamiltonian constraint (\ref{secons1}) and (\ref{secons3}) then gives
\begin{equation}
 \xi B\sqrt{\frac{CR+D}{A\mathcal{Q}+B}} - \Lambda \approx 0\,. \label{eqn:calC_sqrt}
\end{equation}
In order for this weak equation to be non-trivial and to allow for solutions, $\xi B$ and $\Lambda$ must be non-vanishing and have the same sign,
\begin{equation}
 B\ne 0\,, \quad \Lambda\ne 0\,, \quad \xi = \mathrm{sgn} (B\Lambda)\,. \label{eqn:BLambdasign}
\end{equation}
Under the condition (\ref{eqn:BLambdasign}), the Hamiltonian constraint (\ref{eqn:calC_sqrt}) is rewritten as
\begin{equation}
 \mathcal{Q} \approx \frac{B}{A\Lambda^2}(BCR+BD-\Lambda^2)\,. \label{eqn:calQ_sqrt}
\end{equation}
Upon using (\ref{eqn:calQ_sqrt}), the consistency condition (\ref{eqn:dFdt=0_sqrt}) is rewritten as
\begin{eqnarray}\label{cons2}
 0 \approx -\frac{\partial \mathcal{F}}{\partial t}
   \approx
  \dot{\Lambda}+\frac{\Lambda}{2}\left(\frac{\dot{A}}{A}-\frac{\dot{B}}{B}\right)-\frac{B}{2A\Lambda}(\dot{A}D+A\dot{D})-\frac{B}{2A\Lambda}(\dot{A}C+A\dot{C})R\,,
\end{eqnarray}
where we have used the condition (\ref{eqn:BLambdasign}). The above consistency condition (\ref{cons2}) is automatically satisfied and the extended Hamiltonian constraint $\mathcal{C}^E$ is first class if
\begin{eqnarray}
\dot{\Lambda}+\frac{\Lambda}{2}\left(\frac{\dot{A}}{A}-\frac{\dot{B}}{B}\right)-\frac{B}{2A\Lambda}(\dot{A}D+A\dot{D}) & = & 0,\label{cons11}\\
\dot{A}C+A\dot{C} & = & 0.\label{cons22}
\end{eqnarray}
The conditions (\ref{cons11}) and (\ref{cons22}), respectively, give
\begin{eqnarray}
A = \frac{\text{constant}\cdot B}{\Lambda^2-BD},\qquad AC=\text{constant}.
\end{eqnarray}
For example, if $B\propto A\Lambda^2$, $C\propto 1/A$, $D\propto C$ and $\Lambda\ne 0$ then the conditions (\ref{cons11}) and (\ref{cons22}) are satisfied but all explicit time dependence in this case can be absorbed by a redefinition of the lapse function. On the other hand, if $A=C=1$ and $B=\Lambda^2/(D+\mbox{constant})$ for example, then the conditions (\ref{cons11}) and (\ref{cons22}) are satisfied by any function $D(t)$ but the time dependence of coefficients cannot be removed by a simple redefinition of the lapse function.

If (\ref{cons22}) is satisfied but (\ref{cons11}) is not satisfied, then the theory is inconsistent since the consistency condition (\ref{cons2}) is violated. Finally, if $AC\neq \text{constant}$, then the consistency condition (\ref{cons2})  gives a tertiary constraint, which we call $\mathcal{C}_3$. We can check that in this case the Hamiltonian constraint $\mathcal{C}$ and tertiary constraint $\mathcal{C}_3$ are second class, and the rest of the constraints are all first class. There are only 2 local physial degrees of freedom in this case of the theory.

When $BD>0$, it is illustrative to rewrite the action (after integrating out the auxiliary tensor fields) as
\begin{equation}\label{theory1}
S=\int d^4x\sqrt{h}N\left[\xi M(t)^4\sqrt{\left(1+\frac{c_1(t)}{M(t)^2}\mathcal{K}\right)\left(1+\frac{c_2(t)}{M(t)^2}R\right)}-\Lambda(t)\right]\,,
\end{equation}
where $\mathcal{K}=K_{ij}K^{ij}-K^2$, $K=K^i_{\ i}$, $\xi=\pm1$, $M=(BD)^{1/8}$, $c_1=M^2A/B$ and $c_2=M^2C/D$. In the weak gravitational field limit, we expand the action as
\begin{equation}\
S\simeq\int d^4x\sqrt{h}N\left[\xi M^4-\Lambda+\frac{\xi}{2}M^2(c_1\mathcal{K}+c_2R)+...\right]\,. 
\end{equation}
We then demand that $\xi M^4-\Lambda\simeq 0$ to cancel out the bare cosmological constant since we do not intend to address the cosmological constant problem in the present work. The effective Planck scale $M_p$, the sound speed of gravitational waves $c_g$ and the effective cosmological constant $\Lambda_{\rm eff}$ read
\begin{eqnarray}
M_p^2=\xi c_1 M^2\,,\quad c_g^2=\frac{c_2}{c_1}\,,\quad \Lambda_{\rm eff} = \frac{\Lambda-\xi M^4}{\xi c_1M^2}\,,
\end{eqnarray}
where $\xi c_1>0$ and $c_2/c_1>0$ are required to ensure the absence of ghost and gradient instability in the gravity sector.

Now let us investigate the flat FLRW solution in the theory with constant coefficients. In the matter sector, we introduce a canonical scalar field $\phi$ minimally couples to gravity. We take the flat FLRW ansatz,
\begin{equation}
 N = N(t)\,,\quad N^i = 0\,, \quad h_{ij} = a(t)^2\delta_{ij}\,,
\end{equation}
and set $\xi=1$. The action is then reduced to 
\begin{equation}
 S = \int dx^3\int dt a^3\left[M^4\sqrt{N^2-\frac{6c_1}{M^2}\frac{\dot{a}^2}{a^2}}-N\Lambda+\frac{1}{2N}\dot{\phi}^2-NV(\phi)\right]\,.
\end{equation}
Taking the variation of the mini-superspace action with respect to $N$ and $a$, we obtain the Friedmann equations of the form
\begin{eqnarray}
1-6c_1\frac{H^2}{M^2}&=&\frac{M^8}{\left(\Lambda+\rho_m\right)^2},\label{frd1}\\
-\frac{c_1\dot{H}}{NM^2}&=&\frac{M^{8}\dot{\phi}^2}{2N^2\left(\Lambda+\rho_m\right)^3}, \label{frd2}
\end{eqnarray}
where $H=\dot{a}/(Na)$ is the Hubble expansion rate, and $\rho_m=\frac{1}{2N^2}\dot{\phi}^2+V(\phi)$ is the energy density of the scalar field. It is easy to check that the eq. (\ref{frd2}) is consistent with eq. (\ref{frd1}) and thus the Bianchi identity holds, provided that the scalar field $\phi$ satisfies the equation of motion, 
\begin{equation}
 \frac{1}{N}\partial_t\left(\frac{\dot{\phi}}{N}\right) + 3H\frac{\dot{\phi}}{N} + V'(\phi) = 0,
\end{equation}
which follows from the variation of the mini-superspace action with respect to $\phi(t)$. In the limit $\rho_m \to\infty$, we have 
\begin{eqnarray}
H^2\to\frac{1}{6c_1^2}M_p^2,\qquad\text{where}\qquad M_p^2\equiv c_1M^2.
\end{eqnarray}
The Hubble scale could be much smaller than Planck scale if $c_1\gg1$. Thus our theory is free from cosmological singularity, as far as the strictly homogeneous, isotropic and flat universe is concerned. It would be very interesting to investigate the stability of the FLRW solution against inhomogeneous perturbations and applications to the early universe, for instance inflation. However it is beyond the scope of our current paper and thus we would like to defer it to future work. At low energy scale, the eq. (\ref{frd1}) can be approximated as 
\begin{eqnarray}
6c_1M^2H^2=2\rho_m-\frac{3\rho_m^2}{M^4}+...
\end{eqnarray}
we have set $\Lambda=M^4$ to cancel out the bare cosmological constant. The theory thus recovers the standard FLRW solution in Einstein gravity when the energy density of the matter sector is sufficiently lower than $M^4$.

\subsection{An exponential gravity}

We have used the momentum constraints to find the theories (\ref{theory0}) and (\ref{theory1}). Actually we can find other type of examples if we use both the momentum constraints and the Hamiltonian constraint. We consider terms such as $e^{R}$ in the action so that its derivative with respect to the Ricci tensor in (\ref{cond2}) is proportional to itself and thus we are able to use the Hamiltonian constraint in an efficient way. We take the following ansatz, 
\begin{eqnarray}
F=f_1(\mathcal{Q})+\exp\left[c_1R+f_2(\mathcal{Q})\right].
\end{eqnarray}
A combination of the Hamiltonian constraint (\ref{secons1}) and (\ref{secons3}) gives 
\begin{eqnarray}
0\approx F-\frac{\partial F}{\partial Q_{ab}}Q^{ab}=\left(f_1-2f_1'\mathcal{Q}\right)+e^{c_1R+f_2}\left(1-2f_2'\mathcal{Q}\right)\,. \label{eqn:C_exp}
\end{eqnarray}
Hence, the derivative of the function $F$ with respect to the 3-dimensional Ricci scalar reads
\begin{eqnarray}\label{pfpr}
\frac{\partial F}{\partial R}=c_1e^{c_1R+f_2(\mathcal{Q})}\approx-c_1\cdot\frac{f_1-2f_1'\mathcal{Q}}{1-2f_2'\mathcal{Q}}.
\end{eqnarray}
The momentum constraint (\ref{secons2}) gives
\begin{eqnarray}\label{expmomc}
 0 & \approx & \nabla_i\left(\frac{\partial F}{\partial Q_{ij}}\right)\nonumber\\
 &=&\left(2f_1'+e^{c_1R+f_2}\cdot2f_2'\right)\nabla_i\left(Q^{ij}-Qh^{ij}\right)+\left(Q^{ij}-Qh^{ij}\right)\nabla_i\left(2f_1'+e^{c_1R+f_2}\cdot2f_2'\right)\,.\nonumber\\
\end{eqnarray}
The self-consistency condition (\ref{cond2}) thus reduces to 
\begin{eqnarray}
\nabla_i\left[\frac{f_1-2f_1'\mathcal{Q}}{1-2f_2'\mathcal{Q}}\left(f_1'-\frac{f_1-2f_1'\mathcal{Q}}{1-2f_2'\mathcal{Q}}f_2'\right)\right]\approx0,
\end{eqnarray}
where we have used (\ref{pfpr}) and (\ref{expmomc}) to simply the self-consistency condition.
One of solutions to the above equation is that $f_1=c_4(t)\mathcal{Q}+\Lambda(t)$ and $f_2=c_3(t)\mathcal{Q}+\ln{c_2}(t)$, 
where
\begin{eqnarray}
c_4=2\Lambda c_3.
\end{eqnarray}
so that (\ref{eqn:C_exp}) reduces to
\begin{equation}
 e^{c_1R+c_3\mathcal{R}}\approx -\frac{\Lambda}{c_2}\,. \label{eqn:C_exp_simple}
\end{equation}
Clearly, this requires that 
\begin{equation}
 c_2\Lambda<0\,. \label{eqn:c2Lambda<0}
\end{equation}
The extended Hamiltonian constraint $\mathcal{C}^E$ defined by (\ref{eqn:def-CE}), which is a linear combination of $\mathcal{C}$, $P^{ij}$, $U_{ij}$ and $\Psi^{ij}$, is first class if all coefficients in the theory are constant.

If these coefficients are time dependent, the consistency condition requires that 
\begin{eqnarray}
0 &\approx &\frac{1}{\sqrt{h}}\frac{d \mathcal{C}^E}{dt}\approx\frac{1}{\sqrt{h}}\frac{\partial \mathcal{C}_E}{\partial t}\approx\frac{1}{\sqrt{h}}\frac{\partial \mathcal{C}}{\partial t}= -\frac{\partial }{\partial t}\left(F+v^{ij}Q_{ij}\right) = -\frac{\partial F}{\partial t}\nonumber\\
 &\approx& -\mathcal{Q}c_3\Lambda\partial_t\ln(c_1c_3\Lambda^2)-c_1\Lambda\partial_t\left[\frac{\ln(-\Lambda/c_2)}{c_1}\right]\,,
\end{eqnarray}
where we have used (\ref{eqn:C_exp_simple}) to obtain the last expression. The extended Hamiltonian constraint remains first class if 
\begin{equation}
 c_1c_3\Lambda^2 = \mbox{constant}\,,\quad
  \frac{\ln(-\Lambda/c_2)}{c_1} = \mbox{constant}\,.
\end{equation}
If $c_1c_3\Lambda^2=\mbox{constant}$ but $[\ln(-\Lambda/c_2)]/c_1\ne \mbox{constant}$, the theory is inconsistent. Finally, if $c_1c_3\Lambda^2\ne \mbox{constant}$ then the consistency condition gives a tertiary constraint. In this case both of the extended Hamiltonian constraint and this tertiary constraint are second class.

Now let us write down the action of the exponential gravity (after integrating out the auxiliary tensor fields), and expand it in the weak gravitational field limit,
\begin{eqnarray}
F&=&2\Lambda c_3\mathcal{K}+\Lambda+c_2\exp\left[c_1R+c_3\mathcal{K}\right]\\
&=&\Lambda+c_2+\left(2\Lambda c_3+c_2c_3\right)\mathcal{K}+c_1c_2R+\frac{1}{2}c_2(c_1R+c_3\mathcal{K})^2+\cdots\,.
\end{eqnarray}
We then demand $\Lambda+c_2\simeq 0$ to cancel out the bare cosmological constant. General relativity is recovered at low energy limit if all coefficients are constant and satisfy $2\Lambda c_3+c_2c_3>0$ and $c_1c_2>0$.

\subsection{Lapse independent term}

The theories that we just found in previous subsections can be extended to
\begin{eqnarray}\label{genst2}
S=\int dt d^3x\sqrt{h}\left[N F+ G\left(R_{ij},\nabla_i,h^{ij}, t\right)\right],
\end{eqnarray}
where $G$ is a generic spatial scalar made of its arguments, and $F$ is an action that satisfies the self-consistency condition (\ref{cond1}), for instance the Einstein gravity, the square root gravity, as well as the exponential gravity. The addtional term $G(R_{ij},\nabla_i,h^{ij}, t)$ contributes to the Hamiltonian but does not contribute to the primary and secondary constraints as well as the Poisson brackets among them. The consistency condition of the extended Hamiltonian constraint generically gives rise to a tertiary constraint. Generically, the extended Hamiltonian constraint and this new tertiary constraint do not commute. Therefore both of them are second class. The scalar graviton is eliminated by these two second class constraints. At the end we have only two local physical degrees of freedom in the gravity sector. One of examples of this kind in the literature is the Cuscuton scalar field theory \cite{Afshordi:2006ad}\cite{Afshordi:2007yx}\cite{Gomes:2017tzd}. In the following, we present another example with Ricci tensor included in the function $G$, as well as its Hamiltonian analysis.

Let us consider the Einstein gravity modified by the additional term of the form $G=(c_0+c_1R^{ij}R_{ij}+c_2R^2)/2$, without the auxiliary tensors $Q_{ij}$ and $v^{ij}$, 
\begin{eqnarray}
S=\frac{1}{2}\int d^4x\sqrt{h}N\left(K^{ij}K_{ij}-K^2+R\right)+\sqrt{h}\left(c_0+c_1R^{ij}R_{ij}+c_2R^2\right)\,.
\end{eqnarray}
The momenta conjugate to $N$, $N^i$ and $h_{ij}$, respectively, are
\begin{eqnarray}
 \pi_N=\frac{\partial\mathcal{L}}{\partial\dot{N}}=0\,,\quad
  \pi_i=\frac{\partial\mathcal{L}}{\partial\dot{N}^{i}}=0\,,\quad
  \pi^{ij}=\frac{\partial\mathcal{L}}{\partial \dot{h}_{ij}}=\frac{1}{2}\sqrt{h}\left(K^{ij}-Kh^{ij}\right)\,.
\end{eqnarray} 
The Hamiltonian reads
\begin{eqnarray}
H&=&\int d^3x\left( \pi^{ij}\dot{h}_{ij}-L+\lambda_N\pi_N+\lambda^i\pi_i\right)\nonumber\\
&=&\int d^3x\left[N\mathcal{C}+N^i\mathcal{H}_i+\lambda_N\pi_N+\lambda^i\pi_i-\frac{1}{2}\sqrt{h}\left(c_0+c_1R^{ij}R_{ij}+c_2R^2\right)\right],
\end{eqnarray}
where $\pi_N\approx0$ and $\pi_i\approx0$ are primary constraints, and 
\begin{eqnarray}
\mathcal{C}&\equiv&\frac{2}{\sqrt{h}}\left(\pi^{ij}\pi_{ij}-\frac{1}{2}\pi^2\right)-\frac{1}{2}\sqrt{h}R\,,\\
\mathcal{H}_i&\equiv&-2\nabla_j\left(\frac{\pi_i^{~j}}{\sqrt{h}}\right)\,.
\end{eqnarray}
The consistency of the primary constraints with the time evolution gives
\begin{eqnarray}
0 & \approx & \frac{d\pi_N}{dt}=\{\pi_N,H\}=-\mathcal{C}\,,\\
0 & \approx & \frac{d\pi_i}{dt}=\{\pi_i,H\}=-\mathcal{H}_i\,.
\end{eqnarray}
The momentum constraints $\mathcal{H}_i$ are first class due to the spatial diffeomorphism invariance and its consistency with the time evolution does not lead to new constraints. On the other hand, the consistency of the Hamiltonian constraint with the time evolution gives rise to a tertiary constraint,
\begin{eqnarray}
 0 & \approx & \mathcal{C}_3 \equiv
  \frac{d\mathcal{C}}{dt}=\{\mathcal{C},H\}\nonumber\\
 &\approx&
  2\pi\left(c_1R^{ij}R_{ij}+c_2R^2-c_0\right)-4c_1\pi_{ij}R^{ik}R_{k}^{~j}-4c_2\pi_{ij}R^{ij}R+4c_1\sqrt{h}\nabla_i\nabla_j\left(R^{ik}\frac{\pi^j_{~k}}{\sqrt{h}}\right)\nonumber\\
&&-2c_1\square\left(R^{ij}\frac{\pi_{ij}}{\sqrt{h}}\right)+(2c_1+4c_2)\sqrt{h}\square\left(\frac{\pi}{\sqrt{h}}R\right)+4c_2\sqrt{h}\nabla_i\nabla_j\left(\frac{\pi^{ij}}{\sqrt{h}}R\right)\,.
\end{eqnarray}
It is straightforward to check that the Poisson bracket $\{\bar{\mathcal{C}}[\alpha],\bar{\mathcal{C}}_3[\beta]\}$ does not vanish and thus both of the Hamiltonian constraint $\mathcal{C}\approx0$ and the tertiary constraint $\mathcal{C}_3\approx0$ are second class. We define the total Hamiltonian as
\begin{eqnarray}
H_{\rm tot}=\int d^3x\left[N\mathcal{C}+N^i\mathcal{H}_i+\lambda_N\pi_N+\lambda^i\pi_i+\lambda_3 \mathcal{C}_3-\frac{\sqrt{h}}{2}\left(c_0+c_1R^{ij}R_{ij}+c_2R^2\right)\right]\,.
\end{eqnarray}
The algebra closes here and the consistency of $\mathcal{C}\approx 0$ and $\mathcal{C}_3\approx 0$ with the time evolution simply fix the Lagrange multipliers in front of them.  The 3 degrees in $h_{ij}$ are eliminated by 3 first class momentum constraints, and 1 degree is eliminated by 2 second class constraints $\mathcal{C}\approx0$ and $\mathcal{C}_3 \approx0$. We thus conclude that there are only 2 local physical degrees of freedom in the gravity sector.

\section{Conclusion and Discussion}\label{conclusion} 

Searching for theories in which all constraints are first class is an interesting problem. If all renormalizable terms are included and if there is no anomaly, then the structure of such a theory is protected by the gauge symmetries associated with the first class constraints and thus stable against quantum corrections.  In the present work, as a first step we have performed a Hamiltonian analysis for a class of theories whose action is linear in the lapse function and in which the temporal diffeomorphism invariance is broken. We have derived the necessary and sufficient condition for a theory in this class to have two or less local physical degrees of freedom, i.e. (\ref{cond1}). Given this self-consistency condition, one can construct an extended Hamiltonian constraint as prescribed by (\ref{eqn:def-CE}) and (\ref{eqn:consistency-condition-decomposed}). The extended Hamiltonian constraint either is first class or generates a tertiary constraint, depending on whether and how the coefficients in the theory depends explicitly on the time. The scalar graviton associated with the broken temporal diffeomorphism invariance is eliminated by either one first class constraint or two second class constraints. The graviton has thus only two (or less) polarizations.

We have also found that the number of physical degrees of freedom does not change if we include lapse-independent terms to the action. In this case, the consistency of the extended Hamiltonian constraint with the time evolution generically gives rise to a tertiary constraint, and the extended Hamiltonian constraint and the tertiary constraint are second class generically. The scalar graviton in this case is therefore eliminated by these two second class constraints.

Besides the Einstein gravity, we have found several simple modified gravity theories with 2 local physical degrees of freedom, i.e. the one with square-root type action, the one with exponential type action and the Einstein gravity modified by additional lapse-independent terms. It would be intriguing to investigate solar system constraints, cosmological implications, compact objects and so on in those theories.

\section*{Acknowledgments}
We would like to thank N. Afshordi, A. De Felice, M. Sasaki and M. C. Werner for the useful discussions. CL is supported by JSPS postdoc fellowship for overseas researchers, and by JSPS Grant-in-Aid for Scientific Research No. 15F15321. The work of SM is supported by JSPS Grant-in-Aid for Scientific Research No. 17H02890, No. 17H06359, No. 17H06357, and by World Premier International Research Center Initiative (WPI), MEXT, Japan.

\appendix

 \section{Theories with mixed derivative terms}\label{mixder}
If the theory contains mixed derivative terms such as $\nabla_iQ_{jk}\nabla^{i}Q^{jk}$, we have to replace (\ref{eqn:defPhiij}) with 
\begin{eqnarray}
 \Phi^{ij} \equiv \sqrt{h}
  \left\{
   \frac{1}{N\sqrt{h}}\frac{\delta}{\delta Q_{ij}}\bar{F}[N\sqrt{h}]
   + v^{ij} \right\}\,,
\end{eqnarray}
and thus $\Phi^{ij}$ now depends on spatial derivatives of $\ln N$, while other constraints are the same as the ones in section \ref{sccond}. Therefore, we have 
\begin{eqnarray}\label{13}
\{\bar{\pi}_{N}[\lambda],\bar{\Phi}^{ij}[\phi_{ij}]\}\neq0,\qquad \{\bar{P}^{ij}[\chi_{ij}], \bar{\mathcal{C}}[\alpha]\}\neq0\,,
\end{eqnarray}
and do not vanish weakly. The matrix (\ref{bigM}) should now be extended to 
\begin{eqnarray}
\tilde{M}_{\tilde{a}\tilde{b}}(x,y)\equiv\{\tilde{\phi}_{\tilde{a}}(x),\tilde{\phi}_{\tilde{b}}(y)\}\approx
\left(\begin{array}{cccccc}
0& 0 &0_{6}^{T} & 0_{6}^{T}& \hat{u}^T_3& 0_{6}^{T}\\
0& 0 & \hat{u}_{4}^{T} & u_{1}^{T} & 0_{6}^{T} & \hat{u}_{2}^{T}\\
0_{6}&\hat{u}_{4} & 0_{6,6} & 0_{6,6} & A_{1} & 0_{6,6}\\
0_{6}&-u_{1} & 0_{6,6} & 0_{6,6} & a\mathbf{1}_{6,6} & b\mathbf{1}_{6,6}\\
\hat{u}_3&0_{6} & -A_{1}^{T} & -a\mathbf{1}_{6,6} & 0_{6,6} & \hat{A}_{2}\\
0_{6}&-\hat{u}_{2} & 0_{6,6} & -b\mathbf{1}_{6,6} & -\hat{A}_{2} & A_{3}
\end{array}\right)\,,
\end{eqnarray}
where $\tilde{\phi}_{\tilde{a}}\equiv\left(\pi_N, \mathcal{C},P^{ij}, U_{ij},\Phi^{ij},\Psi^{ij}\right)$, $\tilde{a}=1,\cdots,26$. Therefore the structure of the theory with mixed derivative terms can be quite different from that without mixed derivative terms.

\section{BSW action of general relativity}\label{bsw}

General relativity can be interpreted as a theory of evolving 3-geometry. Let's start from general relativity, and derive the Baierlein-Sharp-Wheeler (BSW) action \cite{shape1}. The Einstein-Hilbert action reads
\begin{eqnarray}
S=\int d^4x \sqrt{h}\left[\frac{1}{N}\left(E_{ij}E^{ij}-E^2\right)+N\left(R+\Lambda\right)\right]+\mbox{boundary terms}\,,
\end{eqnarray}
where 
\begin{eqnarray}
E_{ij}\equiv NK_{ij}=\frac{1}{2}\left(\partial_t{h}_{ij}-\nabla_iN_j-\nabla_jN_i\right).
\end{eqnarray}
Taking the variation of the Einstein-Hilbert action with respect to the lapse function $N$, we obtain
\begin{eqnarray}
N=\sqrt{\frac{E_{ij}E^{ij}-E^2}{R+\Lambda}}.
\end{eqnarray}
Plugging this back into the Einstein-Hilbert action, we obtain the BSW action,
\begin{eqnarray}
S=2\int d^4x\sqrt{h}\sqrt{\left(E_{ij}E^{ij}-E^2\right)\cdot \left(R+\Lambda\right)} + \mbox{boundary terms}\,.
\end{eqnarray}

\section{Square root gravity without $Q_{ij}$ and $v^{ij}$}\label{apsqq}

The action of the square root gravity studied in subsection~\ref{subsec:squareroot} is written as  
\begin{eqnarray}
 S=\int d^4x \mathcal{L}\,, \quad
  \mathcal{L} = \sqrt{h}N\left[\xi \sqrt{\left(A\mathcal{K}+B\right)\left(C R+D\right)}-\Lambda\right],
\end{eqnarray}
where $\mathcal{K}\equiv K_{ij}K^{ij}-K^2$, $\xi=\pm1$ and all coefficients $A, B, C, D$ and $\Lambda$ are some functions of time. We will perform the Hamiltonian analysis of this theory without introducing the auxiliary tensor fields $Q_{ij}$ and $v^{ij}$. The conjugate momenta reads
\begin{eqnarray}
 \pi_N=\frac{\partial\mathcal{L}}{\partial\dot{N}}=0\,,\quad \pi_i=\frac{\partial\mathcal{L}}{\partial\dot{N}_{i}}=0\,,\quad
\pi^{ij}=\frac{\partial\mathcal{L}}{\partial\dot{h}_{ij}}=\frac{\xi\sqrt{h}A}{2} \sqrt{\frac{C R +D}{A\mathcal{K}+B}}\left(K^{ij}-Kh^{ij}\right)\,. 
\end{eqnarray}
The Hamiltonian reads
\begin{eqnarray}
H=\int d^3x \left[\lambda_c\mathcal{C}+\tilde{N}^i\mathcal{H}_i+\lambda_N\pi_N+\lambda_i\pi^i\right]\,,
\end{eqnarray}
where $\lambda_c$, $\tilde{N}^i$, $\lambda_N$, $\lambda_i$ are Lagrange multipliers and 
\begin{eqnarray}
\mathcal{C}&\equiv&-\xi \sqrt{h}B\sqrt{\frac{CR+D}{A\mathcal{K}+B}}+\sqrt{h}\Lambda\nonumber\\
&=& -\xi \sqrt{h}B^{1/2}\left[C R+D-\frac{4\left(\pi_{ij}\pi^{ij}-\frac{1}{2}\pi^2\right)}{A\cdot h}\right]^{1/2}+\sqrt{h}\Lambda\,,\label{srgHam}\\
\mathcal{H}_i&\equiv&-2\sqrt{h}\nabla_j\left(\frac{\pi^i_{~i}}{\sqrt{h}}\right)\,. 
\end{eqnarray}
We have 8 constraints: the Hamiltonian constraint $\mathcal{C}\approx 0$, the momentum constraints $\mathcal{H}_i\approx 0$ as well as the primary constraints $\pi_N\approx 0$ and $\pi_i\approx 0$.

If the coefficients $A, B, C, D$ and $\Lambda$ are constants, all Poisson brackets between any pair of constraints weakly vanish:  
\begin{eqnarray}
\{\bar{\mathcal{C}}[\alpha],\bar{\mathcal{C}}[\beta]\}&=&\int d^3x \frac{B^2C}{A}\left[\frac{\beta\sqrt{h}}{\sqrt{h}\Lambda-\mathcal{C}}\cdot\nabla^i\left(\frac{\alpha\sqrt{h}}{\sqrt{h}\Lambda-\mathcal{C}}\right)-\left(\alpha\leftrightarrow\beta\right)\right]\mathcal{H}_i\approx0,\\
\{\bar{\mathcal{C}}[\alpha],\bar{\mathcal{H}}_i[f^i]\}&=&\int d^3x \frac{1}{2}\left(\alpha\nabla_if^i-f^i\nabla_i\alpha\right)\mathcal{C}\approx0,
\end{eqnarray}
and the Poisson brackets of constraints $\pi_N$ and $\pi_i$ with any constraints all vanish strongly. Therefore, all 8 constraints are first class and the algebra closes here. There are only 2 local physical degrees of freedom in the theory.


\begin{thebibliography}{99}

\bibitem{Horava:2009uw} 
  P.~Horava,
  Phys.\ Rev.\ D {\bf 79}, 084008 (2009)
  [arXiv:0901.3775 [hep-th]].

\bibitem{Mukohyama:2010xz} 
  S.~Mukohyama,
  Class.\ Quant.\ Grav.\  {\bf 27}, 223101 (2010)
  [arXiv:1007.5199 [hep-th]].

\bibitem{Barvinsky:2015kil} 
  A.~O.~Barvinsky, D.~Blas, M.~Herrero-Valea, S.~M.~Sibiryakov and C.~F.~Steinwachs,
  Phys.\ Rev.\ D {\bf 93}, no. 6, 064022 (2016)
  [arXiv:1512.02250 [hep-th]].

\bibitem{Ratra:1987rm} 
  B.~Ratra and P.~J.~E.~Peebles,
  Phys.\ Rev.\ D {\bf 37}, 3406 (1988).
 
 \bibitem{Caldwell:1997ii} 
  R.~R.~Caldwell, R.~Dave and P.~J.~Steinhardt,
  Phys.\ Rev.\ Lett.\  {\bf 80}, 1582 (1998)
  [astro-ph/9708069].
  
\bibitem{ArkaniHamed:2003uy} 
  N.~Arkani-Hamed, H.~C.~Cheng, M.~A.~Luty and S.~Mukohyama,
  JHEP {\bf 0405}, 074 (2004)
  [hep-th/0312099].

\bibitem{FPtheory}
 M. Fierz, W. Pauli, Proc. Roy. Soc. Lond. A173, 211-232 (1939).

\bibitem{deRham:2010kj} 
  C.~de Rham, G.~Gabadadze and A.~J.~Tolley,
  Phys.\ Rev.\ Lett.\  {\bf 106}, 231101 (2011)
  [arXiv:1011.1232 [hep-th]].

\bibitem{Gumrukcuoglu:2011ew} 
  A.~E.~Gumrukcuoglu, C.~Lin and S.~Mukohyama,
  JCAP {\bf 1111}, 030 (2011)
  [arXiv:1109.3845 [hep-th]].

\bibitem{Ostrogradsky}
M. Ostrogradskyy: Mem. Ac. St. Petersbourg VI 4, 385 (1850).

\bibitem{Woodard:2006nt} 
  R.~P.~Woodard,
  Lect.\ Notes Phys.\  {\bf 720}, 403 (2007)
  [astro-ph/0601672].

\bibitem{Horndeski}
 G. W. Horndeski, Int. J. Theor. Phys. 10, 363 (1974).


\bibitem{Nicolis:2008in} 
  A.~Nicolis, R.~Rattazzi and E.~Trincherini,
  Phys.\ Rev.\ D {\bf 79}, 064036 (2009)
  [arXiv:0811.2197 [hep-th]].

\bibitem{Deffayet:2009wt} 
  C.~Deffayet, G.~Esposito-Farese and A.~Vikman,
  Phys.\ Rev.\ D {\bf 79}, 084003 (2009)
  [arXiv:0901.1314 [hep-th]].

\bibitem{Deffayet:2011gz} 
  C.~Deffayet, X.~Gao, D.~A.~Steer and G.~Zahariade,
  Phys.\ Rev.\ D {\bf 84}, 064039 (2011)
  [arXiv:1103.3260 [hep-th]].

\bibitem{Zumalacarregui:2013pma} 
  M.~Zumalac\'arregui and J.~Garc\'ia-Bellido,
  Phys.\ Rev.\ D {\bf 89}, 064046 (2014)
  [arXiv:1308.4685 [gr-qc]].

\bibitem{Gleyzes:2014dya} 
  J.~Gleyzes, D.~Langlois, F.~Piazza and F.~Vernizzi,
  Phys.\ Rev.\ Lett.\  {\bf 114}, no. 21, 211101 (2015)
  [arXiv:1404.6495 [hep-th]].

\bibitem{Gao:2014soa} 
  X.~Gao,
  Phys.\ Rev.\ D {\bf 90}, 081501 (2014)
  [arXiv:1406.0822 [gr-qc]].

\bibitem{Gao:2014fra} 
  X.~Gao,
  Phys.\ Rev.\ D {\bf 90}, 104033 (2014)
  [arXiv:1409.6708 [gr-qc]].

\bibitem{Lin:2014jga} 
  C.~Lin, S.~Mukohyama, R.~Namba and R.~Saitou,
  JCAP {\bf 1410}, no. 10, 071 (2014)
  [arXiv:1408.0670 [hep-th]].

\bibitem{Langlois:2015cwa} 
  D.~Langlois and K.~Noui,
  JCAP {\bf 1602}, no. 02, 034 (2016)
  [arXiv:1510.06930 [gr-qc]].

\bibitem{Langlois:2015skt} 
  D.~Langlois and K.~Noui,
  JCAP {\bf 1607}, no. 07, 016 (2016)
  [arXiv:1512.06820 [gr-qc]].

\bibitem{BenAchour:2016fzp} 
  J.~Ben Achour, M.~Crisostomi, K.~Koyama, D.~Langlois, K.~Noui and G.~Tasinato,
  JHEP {\bf 1612}, 100 (2016)
  [arXiv:1608.08135 [hep-th]].

\bibitem{lovelock1}
 D. Lovelock, J. Math. Phys. 12, 498 (1971).

\bibitem{lovelock2}
 D. Lovelock, J. Math. Phys. 13, 874 (1972)

\bibitem{Saitou:2016lvb} 
  R.~Saitou,
  Phys.\ Rev.\ D {\bf 94}, no. 10, 104054 (2016)
  [arXiv:1604.03847 [hep-th]].

\bibitem{Lin:2017utd} 
  C.~Lin,
  arXiv:1702.00696 [gr-qc].

 \bibitem{Henneaux:2009zb}
  M.~Henneaux, A.~Kleinschmidt and G.~Lucena Gómez,
  Phys.\ Rev.\ D {\bf 81} (2010) 064002
  [arXiv:0912.0399 [hep-th]].

\bibitem{Blas:2009yd} 
  D.~Blas, O.~Pujolas and S.~Sibiryakov,
  JHEP {\bf 0910}, 029 (2009)
  [arXiv:0906.3046 [hep-th]].

\bibitem{shape1}
 R.F. Baierlein, D. Sharp and J.A. Wheeler, Phys. Rev. 126 1864 (1962).

\bibitem{Mercati:2014ama} 
  F.~Mercati,
  arXiv:1409.0105 [gr-qc].

\bibitem{Murchadha:2012zz} 
  N.~O.~Murchadha, C.~Soo and H.~L.~Yu,
  Class.\ Quant.\ Grav.\  {\bf 30}, 095016 (2013)
  [arXiv:1208.2525 [gr-qc]].
  
\bibitem{Afshordi:2006ad} 
  N.~Afshordi, D.~J.~H.~Chung and G.~Geshnizjani,
  Phys.\ Rev.\ D {\bf 75}, 083513 (2007)
  [hep-th/0609150].
 
\bibitem{Afshordi:2007yx} 
  N.~Afshordi, D.~J.~H.~Chung, M.~Doran and G.~Geshnizjani,
  Phys.\ Rev.\ D {\bf 75}, 123509 (2007)
  [astro-ph/0702002].

\bibitem{Gomes:2017tzd} 
  H.~Gomes and D.~C.~Guariento,
  Phys.\ Rev.\ D {\bf 95}, no. 10, 104049 (2017)
  [arXiv:1703.08226 [gr-qc]].
 
\end{thebibliography}
\end{document}